\documentclass[reprint,amsmath,amssymb,aps,pre]{revtex4-1}
\usepackage{bm}
\usepackage{mathtools}
\usepackage[dvips]{graphicx}

\begin{document}
\title{
  Critical Nonequilibrium Cluster-flip Relaxations in Ising Models
}
\author{Yusuke Tomita}
\affiliation{College of Engineering, Shibaura Institute of Technology, Saitama, Saitama 337-8570, Japan}
\author{Yoshihiko Nonomura}
\affiliation{International Center for Materials Nanoarchitectonics, National Institute for Materials Science, Tsukuba, Ibaraki 305-0044, Japan}

\date{\today}

\begin{abstract}
  We investigate nonequilibrium relaxations of Ising models at the critical
  point by using a cluster update.
  While preceding studies imply that nonequilibrium cluster-flip dynamics
  at the critical point are universally described
  by the stretched-exponential function,
  we find that the dynamics changes from the stretched-exponential
  to the power function as the dimensionality is increased:
  The two-, three-, four-, and infinite-dimensional Ising models are
  numerically studied, and the four- and infinite-dimensional
  Ising models exhibit the power-law relaxation.
  We also show that
  the finite-size scaling analysis using the normalized correlation length
  is markedly effective for the analysis of relaxational processes
  rather than the direct use of the Monte Carlo step.
\end{abstract}

\maketitle

\section{%
  \label{sec:intro}
  Introduction
}

Studies of relaxational processes do not merely give
insights but also give probes to investigate
critical phenomena.
The droplet theory~\cite{HuseFisher1987,Tang1989} gives relations between relaxational processes
and droplet excitations, and these enable us to study
droplet structures by observing relaxations of order parameters~\cite{Tomita2016}.
The nonequilibrium relaxation (NER) method gives us
an alternative mean to investigate critical phenomena~\cite{OzekiIto2007}.
The NER functions show power-law relaxations at a critical point,
whereas they are exponential at off-critical points.
The powers of the power functions consist of critical
and dynamical exponents.
The NER method estimates a critical point by finding a point
where relaxations exhibit power-law relaxations,
and critical exponents are estimated by applying a scaling ansatz
to the power-law relaxation functions.
While the NER method examines relaxational processes
of sudden cooling and/or heating,
a method using Kibble-Zurek scaling ansatz
examines relaxational processes of scheduled cooling and/or heating~\cite{Kibble1976,Zurek1985,Chandran2012,Liu2014,Xu2018}.
The way of using scheduled temperature changing protocol
enables us to study a variety of nonequilibrium relaxations
in spin systems.

Although examining dynamical quantities rather than static ones
is an indirect approach,
it is effective for systems whose relaxations are slow.
Methods utilizing general relations
in the nonequilibrium relaxation do not require thermalization;
all the production runs are started
without discarding Monte Carlo steps for thermalization.

Since cluster-flip update algorithms~\cite{SwendsenWang1987,Wolff1989}
significantly accelerate Monte Carlo simulation,
the algorithms are widely used especially for extensive simulations~\cite{KomuraOkabe2012XY}.
The algorithms unite correlated spins into a cluster
and flip clusters at a time.
The global update reduces autocorrelation time
and enables us to sample evenly in a state space with small Monte Carlo steps.

One of the authors proposed a new method which
integrates the NER method and the cluster-flip update.
In the preceding study, it was found that
the nonequilibrium relaxation of the Ising model at the critical point
is not described by the power-law, which is expected
in the NER method with single-spin-flip updates,
but by the stretched-exponential relaxations~\cite{Nonomura2014}.
Our consecutive studies revealed that the stretched-exponential relaxations
are common in the NER with the cluster-flip update~\cite{NonomuraTomita2015,NonomuraTomita2016}.

To further develop
the method which integrates the NER method and cluster-flip update,
an understanding of the origin of the stretched-exponential relaxation
is indispensable.
In this paper, critical nonequilibrium relaxations with cluster-update in
two-, three-, four-, and infinite-dimensional
Ising models are examined.
Through the systematical change in the dimensionality,
the origin of the stretched-exponential relaxation
is investigated.

The remains of the paper are organized as follows.
In Sec.~\ref{sec:model} the Hamiltonian and
the numerical method used in the paper are given.
Results for several dimensional Ising models are given
in Sec.~\ref{sec:result}.
The analysis used in the paper can be extended to a general procedure
for investigation of critical phenomena,
and the procedure is summarized in Sec.~\ref{sec:cner}.
Section~\ref{sec:discussion} is devoted to summary and discussion.

\section{%
  \label{sec:model}
  Model and Method
}

We investigate the NER at critical points for
two-, three-, four-, and infinite-dimensional Ising models.
The Hamiltonian of the finite-dimensional Ising model is given by
\begin{equation}
  \mathcal{H} = -J\sum_{\langle i, j\rangle}\sigma_i\sigma_j,
\end{equation}
where $J$ is the exchange coupling constant,
$\sigma_i(\in \{\pm 1\})$ is an Ising spin at site $i$,
and the sum runs over nearest neighbors.
On the other hand,
the Hamiltonian of the infinite-range Ising model is given by
\begin{equation}
  \label{eq:irmodel}
  \mathcal{H} = -\frac{J}{N}\sum_{i < j}\sigma_i\sigma_j,
\end{equation}
where $N$ is the number of sites.
The normalization factor $N$ is required to make the system extensive.
Hereafter, the Boltzmann constant and the exchange coupling constant $J$
are set to unity.

In the present paper,
we observe NERs from totally random states ($T=\infty$)
to critical states ($T=T_{\mathrm{c}}$) in the Ising models
using the cluster update.
In order to update the systems uniformly,
we employ the Swendsen-Wang (multiple) cluster update~\cite{SwendsenWang1987,KomuraOkabe2012, Komura2015}.
The Wolff (single) cluster update~\cite{Wolff1989} tends to flip
larger clusters, and it could bring about spatially non-uniform update
especially in the initial stage.

Computational cost of the Swendsen-Wang cluster algorithm
for the infinite-range model is $O(N^2)$
since it scans all the $N^2$ bonds.
To reduce the cost to $O(N)$,
the $O(N)$ cluster Monte Carlo method~\cite{FukuiTodo2009} is employed.
The method skips the bond scan properly
and realizes $O(N)$ computational cost.

\section{%
  \label{sec:result}
  Results
}

\subsection{Infinite-range Ising model}

In order to understand the NER in the cluster-flip update,
we consider time evolution in the infinite-range (IR) Ising model first
since the model is one of the simplest models which exhibits
phase transitions at a finite temperature.
The IR Ising model is an Ising model on the complete graph.
Owing to the special character of the complete graph,
the development of the magnetization from the completely random
configuration ($T=\infty$) to the critical state
can be described by the simple exponential form (see Appendix~\ref{appendix:b}).
We refine the exponential form (Eq.~(\ref{eq:mn})) by considering three facts:
(i) The magnetization per site converges to its thermal equilibrium states
$\langle m_{\mathrm{C}}\rangle$,
which is given by~(\ref{eq:irmodel_mag_tc}),
\begin{align}
  N^{1/4}\langle m_{\mathrm{C}}\rangle &= \frac{12^{1/4}\Gamma(1/2)}{\Gamma(1/4)},\\
  &\sim 0.909890588\nonumber,
\end{align}
where $\Gamma(x)$ is the gamma function.
(ii) The first step of the time evolution is the percolation process
on the complete graph at the critical point,
and the magnetization is proportional to $N^{\beta_{\mathrm{p}}/d_{\mathrm{p}}\nu_{\mathrm{p}}}$,
where $\beta_{\mathrm{p}}(= 1)$ and $\nu_{\mathrm{p}}(= 1/2)$ are, respectively, the mean-field critical exponents
of the percolation problem for the spontaneous magnetization
and the correlation length.
The upper critical dimension of the percolation problem $d_{\mathrm{p}}$ is six.
(iii) The time scale per one Monte Carlo step depends on the system size.
The growth rate of the correlation in the cluster-flip update
is proportional to the correlation length $\xi$ and merging rate $S(t)$,
and it is described by
\begin{equation}
  \label{eq:diff_xi}
  \frac{d\xi}{dt} = S(t)\xi.
\end{equation}
Owing to the peculiarity of the complete graph,
the merging rate $S(t)$ at the criticality
is proportional to the number of outside sites
of ordered domains times the density of bonds,
\begin{equation}
  \label{eq:mr_irising}
  S(t) = A\frac{M_{\infty} - M(t)}{N},
\end{equation}
where $M_{\infty}$, $M(t)$, and $A$ are, respectively,
the magnetization at the thermodynamic limit,
the magnetization at Monte Carlo step $t$, and
a constant.
Considering that $M \propto N^{3/4}$ (see Appendix A),
Eq.~(\ref{eq:diff_xi}) gives that
the time scale of the IR Ising model is proportional to $N^{-1/4}$.
Taking account of the above three facts (i)-(iii),
the scaling function form of the magnetization
per site at Monte Carlo step $t$ 
is refined as
\begin{equation}
  \label{eq:irising_mag}
  m(t) = \langle m_{\mathrm{C}}\rangle
  [1 - \{1-c_1(t/N^{1/4})^{2/3}\}\exp(-c_2 t/N^{1/4})]^{1/2}.
\end{equation}
Here, $c_1$ and $c_2$ are coefficients.

\begin{figure}[h]
  \includegraphics[width=0.4\textwidth]{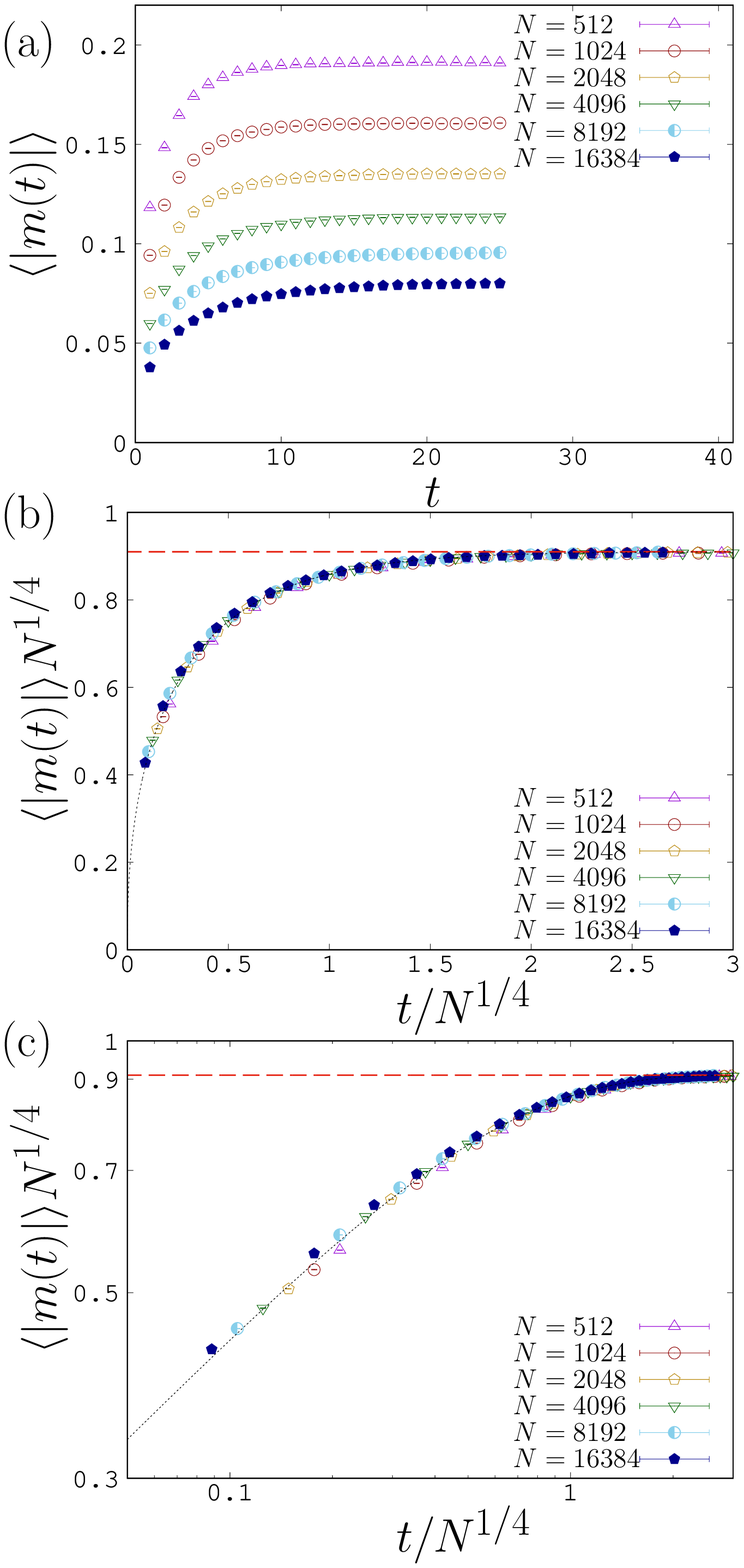}
  \caption{\label{fig:irising}
    (a) Magnetizations per site of IR Ising model as functions of Monte Carlo step $t$
    for several system sizes.
    (b) Scaling plot of the magnetization per site of IR Ising model in a linear scale
    and (c) the same plot in a logarithmic scale.
    The curve shows the fitting result of Eq.~(\ref{eq:irising_mag}).
    The red broken line indicates the thermodynamic limit value,
    $12^{1/4}\Gamma(1/2)/\Gamma(1/4)$.
    Error bars are smaller than the size of the symbols.
  }
\end{figure}

Monte Carlo simulations of the IR Ising model
with the $O(N)$ cluster Monte Carlo method~\cite{FukuiTodo2009}
are executed to obtain the numerical data.
To obtain sample means, $8\times 10^5$ independent runs are executed
for each system size, $N = 1024, 2048, 4096, 8192$, and $16384$.
Figure~\ref{fig:irising}(a) shows magnetizations per site
as functions of Monte Carlo step $t$ for several system sizes,
and the finite-size scaling (FSS) plot is shown in Fig.~\ref{fig:irising}(b)
and (c).
The constants $c_1$ and $c_2$ in Eq.~(\ref{eq:irising_mag})
are $0.463(2)$ and $0.627(2)$, respectively.
The number in parenthesis represents one standard error
in the last digit.
The finite-size scaling function is well described by a single curve,
and it confirms the non-equilibrium relaxation of the cluster-flip update
in the IR Ising model is the product of the power and
simple exponential functions.
At the very beginning of the nonequilibrium relaxation,
the magnetization shows the power-law relaxation,
and it exponentially converges to the thermal equilibrium value
(see Fig.~\ref{fig:irising}(c)).
The Monte Carlo step is scaled by $N^{1/4}(= L)$;
that is, the dynamical exponent is unity
since the effective dimension of the model is four.

\subsection{Finite-dimensional Ising model}

In our previous papers, we observed stretched-exponential relaxations
in the cluster-flip NER at critical points~\cite{Nonomura2014,NonomuraTomita2015,NonomuraTomita2016}.
The observation of the stretched-exponential relaxation
indicates that the cluster-flip NER
is essentially faster than the power-law relaxation which is usually observed in
the single-spin-flip NER.
The essential difference comes from growing processes in ordering.
While the most of spin updates take place on boundaries of
domains of the order parameter in the single-spin-flip update,
domains of the order parameter merge into larger domains
in the cluster-flip update.
This indicates, as explained in the previous subsection (see Eq.~(\ref{eq:diff_xi})),
the relaxational time in the cluster-flip update
is proportional to the correlation length.
The nonequilibrium relaxation process in the finite-dimensional ($d=2, 3,$ and $4$) Ising model can be different from the IR Ising model.
In an ideal situation like the IR Ising model,
a major cluster merges immediate clusters
at each Monte Carlo step,
and the correlation length develops ballistically.
In the finite-dimensional model, however,
a growth rate of clusters fluctuates from place to place,
and relative size of clusters surrounding a major cluster
tends to be smaller as the system evolves.
If we assume that the merging rate in Eq.~(\ref{eq:diff_xi}) as
\begin{equation}
  \label{eq:merging_rate}
  S(t) = ct^{\sigma-1} \quad (\sigma < 1),
\end{equation}
we obtain
\begin{equation}
  \label{eq:xi_strexp}
  \xi(t) = c\exp[(t/\tau)^\sigma],
\end{equation}
where $c$ and $\tau$ are constants.
The stretched-exponential relaxation conforms to
our previous results~\cite{Nonomura2014,NonomuraTomita2015,NonomuraTomita2016}.

\begin{figure}[h]
  \includegraphics[width=0.4\textwidth]{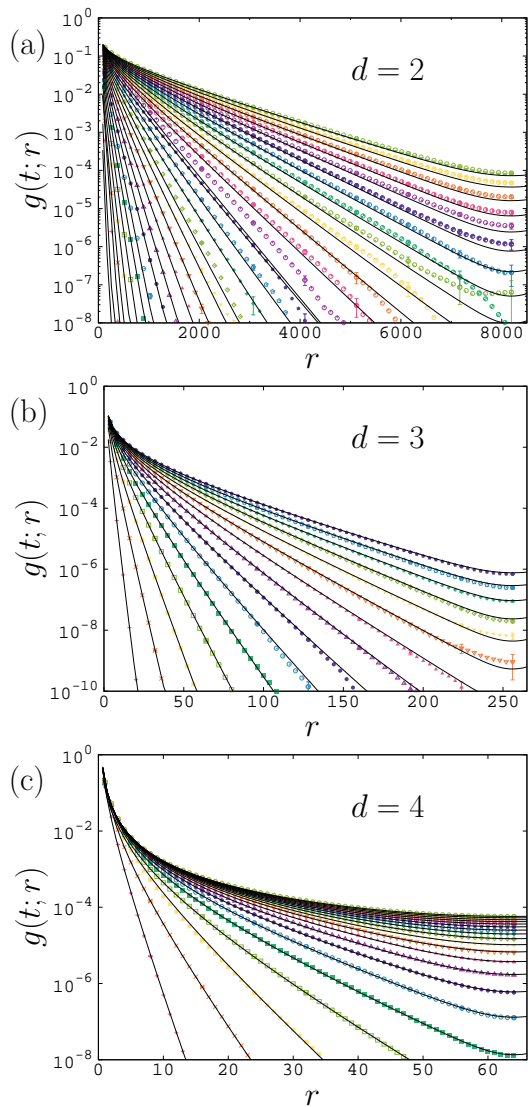}
  \caption{\label{fig:twopfunc}
    Two-point function for (a) the two-, (b) three-, and (c) four-dimensional
    Ising models.
    Curves show the fitting results of Eq.~(\ref{eq:g_fitting_form}).
    Slopes are proportional to the inverse of correlation length.
    As the correlation length develops with time,
    slopes of curves change from steep to gentle.
    To avoid impairing the visibility of figures,
    error bars are plotted for every 8th data point.
  }
\end{figure}

To examine the stretched-exponential development of the correlation length,
we estimate the correlation length as a function of Monte Carlo step $t$,
$\xi(t)$, by the two-point correlation function.
The time-dependent two-point correlation function is given by
\begin{equation}
  g(t; r) = \langle\sigma(t; r_0)\sigma(t; r_0 + r)\rangle,
\end{equation}
where $\sigma(t; r)$ is Ising spin variable at a time $t$ and a position $r$.
To estimate the two-point function effectively,
an improved estimator is used,
\begin{equation}
  g(t; r_i, r_j) = \langle \delta(t; r_i, r_j) \rangle,
\end{equation}
where $\delta(t; r_i, r_j)$ is unity if sites at $r_i$ and $r_j$
belong to the same cluster at a time $t$ and is zero otherwise.
The angle brackets $\langle\cdots\rangle$ denote a sample average.
We estimate the correlation length by assuming
the functional form of the two-point function as
\begin{equation}
  \label{eq:g_fitting_form}
  g(t; r) = A\left(\frac{e^{-r/\xi(t)}}{r^{d-2+\eta}}
  + \frac{e^{-(L-r)/\xi(t)}}{(L-r)^{d-2+\eta}}\right),
\end{equation}
where $r$, $L$, $\eta$, $d$, and $A$ are, respectively,
a spatial distance between two spins, a system size,
the critical exponent of the correlation function,
the spatial dimension, and a constant \footnote{%
  We took into account a correction-to-scaling for
  data of $d=3$ and $L=256$.
  The scaling form we considered is
  $g(r) = A(\exp(-r/\xi)/(r^{d-2+\eta}(1 + ar^\omega))
  + \exp(-(L-r)/\xi)/((L-r)^{d-2+\eta}(1 + a(L-r)^\omega)))$,
  where $\omega$ and $a$ are a correction-to-scaling exponent
  and a constant, respectively.

  Fitting of the four-dimensional Ising model data suffers
  from corrections in small $r$.
  To suppress the corrections, for the four-dimensional data,
  the logarithmic residuals are minimized instead of residuals.
}.
The function form is symmetrized, taking account of
periodic boundary conditions, which are imposed on our Monte Carlo simulations.
Figure~\ref{fig:twopfunc} shows
the two-point functions for the two-, three-, and four-dimensional Ising model.
For the two-dimensional model, the temperature is set
to the critical temperature, $T_{\mathrm{c}} = 2/\ln(1 + \sqrt{2})$.
For the three- and four-dimensional models,
critical temperatures are estimated by durations to reach
equilibrium states in NER,
and they are, respectively, 4.511525 for $d=3$ and 6.680400 for $d=4$.
To obtain sample means, $10^4$ independent runs are executed
for each system size.

\begin{figure}[h]
  \includegraphics[width=0.4\textwidth]{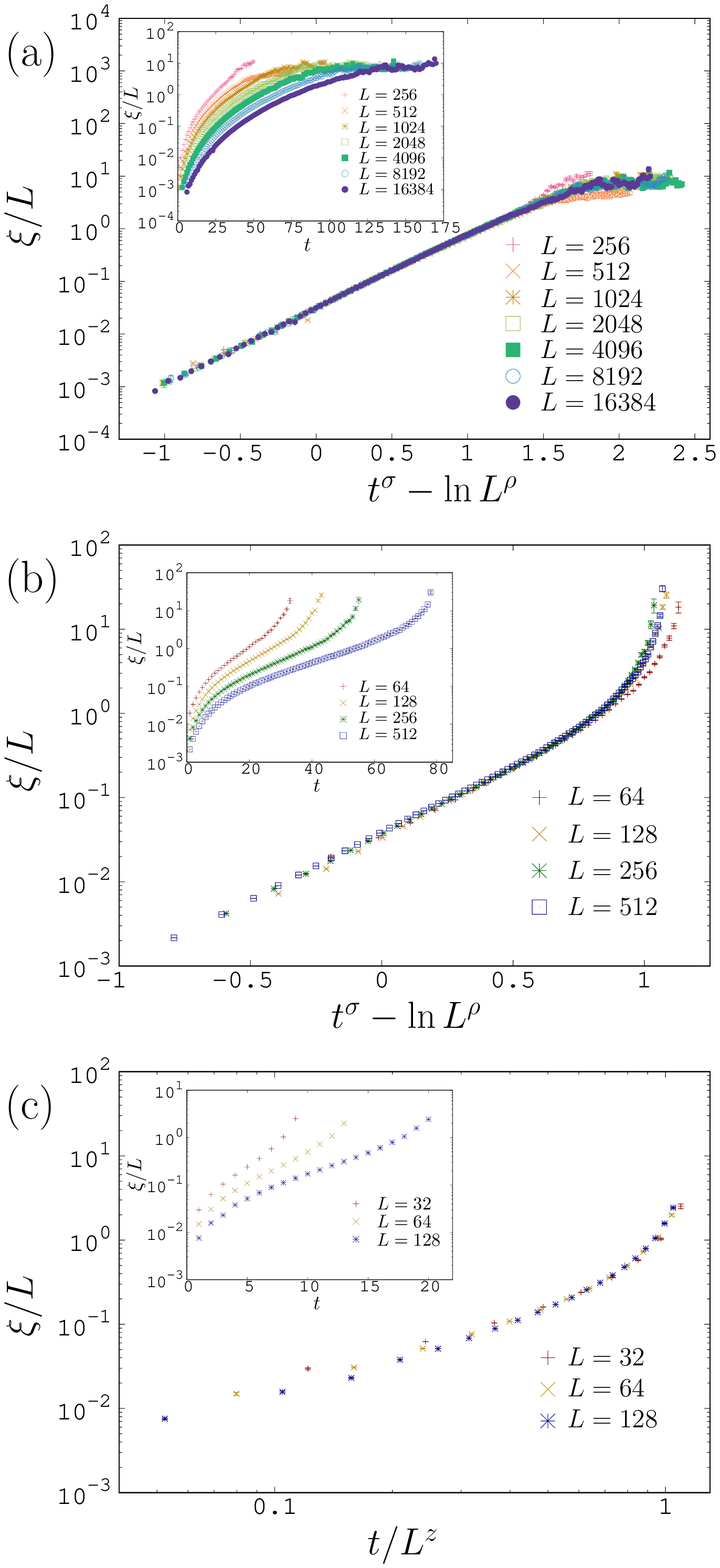}
  \caption{\label{fig:xi}
    Scaling plots of normalized correlation lengths for (a) the two-, (b) three-, and (c) four-dimensional Ising models.
    Insets show the normalized correlation lengths as a function of Monte Carlo
    step $t$.
    Error bars are plotted but are barely visible.
  }
\end{figure}

The correlation lengths estimated from the two-point functions
for the two-, three-, and four-dimensional Ising models
are plotted in insets of Fig.~\ref{fig:xi}.
To examine the development of the correlation length,
we assume the following form for the stretched-exponential relaxation,
\begin{equation}
  \label{eq:xi_form_strexp}
  \xi(t) = A\exp(t^{\sigma}/\rho),
\end{equation}
where $A, \rho$, and $\sigma$ are constants.
Since the dimension of the correlation length is $L$,
we obtain a dimensionless parameter by dividing
the correlation length by the system size $L$,
\begin{equation}
  \label{eq:nxi_form_strexp}
  \xi(t)/L = A\exp[(t^{\sigma} - \rho\ln L)/\rho].
\end{equation}
The normalized correlation length gives
a system-size independent measure of ordering.
The normalized correlation length as a function of $t$
(Eq.~(\ref{eq:nxi_form_strexp})) indicates
the scaling variable of $\xi(t)/L$ is $t^{\sigma} - \ln L^{\rho}$,
and the two- and three-dimensional data are well scaled
by the variable (see Fig.~\ref{fig:xi}(a) and (b)).
The parameters for the FSS plot are, respectively,
$\sigma=0.314(2)$ and $\rho = 0.290(4)$ for the two-dimensional model and
$\sigma=0.241(7)$ and $\rho = 0.29(2)$ for the three-dimensional model.
While the development of the correlation length is described by
the stretched-exponential function for the two- and three-dimensional
Ising models,
the FSS scaling plot using Eq.~(\ref{eq:nxi_form_strexp})
fails for the four-dimensional Ising model.
The failure stems from the smallness of the parameter $\sigma$.
In the limit $\sigma \to 0$, Eqs~(\ref{eq:diff_xi}) and (\ref{eq:merging_rate}) deduce a power-law relaxation,
\begin{equation}
  \label{eq:xi_form_pow}
  \xi(t) = At^\alpha,
\end{equation}
where $A$ and $\alpha$ are constants.
The normalized correlation length for the power-law relaxation
is given by
\begin{equation}
  \label{eq:nxi_form_pow}
  \xi(t)/L = A(t/L^{1/\alpha})^\alpha,
\end{equation}
so that the scaling variable is $tL^{-1/\alpha}$.
Figure~\ref{fig:xi}(c) shows the finite-size scaling plot of
the normalized correlation length for the four-dimensional Ising model.
The dynamical exponent $z(=1/\alpha)$ is estimated as $0.61(1)$.
The upper critical dimension of the Ising model is four,
and the power-law form conforms to the result of the IR Ising model.
However, the dynamical exponent is smaller than
that of the IR Ising model, $z=1$.


In order to see that $\xi(t)/L$ is an essential measure of
the cluster-flip dynamics,
magnetizations per site are plotted as functions of $\xi(t)/L$
in Fig.~\ref{fig:magxi}.
In the usual FSS scaling for equilibrium states,
$(T-T_c)L^{1/\nu}(\propto (L/\xi)^{1/\nu})$ is used as the scaling variable,
and we replace it with $\xi(t)/L$.
The modified FSS expression for the magnetization per site is
\begin{equation}
  \label{eq:fss_mag}
  \langle |m(t)|\rangle = L^{-\beta/\nu}\tilde{m}(\xi(t)/L),
\end{equation}
where $\tilde{m}$ is a scaling function.
In the left-hand side, the thermal average of
the absolute value of the magnetization per site is estimated,
since the sign of the magnetization frequently flips accompanying with a
flip of a major cluster.
The exact critical exponents,
$\beta = 1/8$ and $\nu = 1$ for the two-dimensional
and $\beta = 1/2$ and $\nu = 1/2$ for the four-dimensional Ising model,
are used for the scaling plot for the two- and four-dimensional data.
For the three-dimensional data, $\beta/\nu = 0.5181489(10)$ is used,
which is obtained from the conformal bootstrap with mixed correlators~\cite{Kos2016}.
At the beginning of the nonequilibrium relaxation,
the scaling function $\tilde{m}$
is proportional to the power of $\xi(t)/L$.
Solid lines in Fig.~\ref{fig:magxi}
are fitting results, and their powers are, respectively,
0.8947(4), 0.9423(7), and 0.901(2) for
the two-, three-, and four-dimensional Ising models.

\section{%
  \label{sec:cner}
  Cluster Nonequilibrium Relaxation Method Using Normalized
  Correlation Length
}

The analysis of the nonequilibrium relaxation of the correlation length
in the present paper shows the FSS analysis using the normalized
correlation length, $\xi(t)/L$, is markedly effective.
Although the FSS analysis using $\xi/L$ has been widely used
for the analysis of the thermally equilibrated systems~\cite{Caracciolo1995a,Caracciolo1995b,Tomita2014},
it is worthwhile to describe a procedure of the FSS analysis for
the nonequilibrium relaxation.

\begin{figure}[h]
  \includegraphics[width=0.4\textwidth]{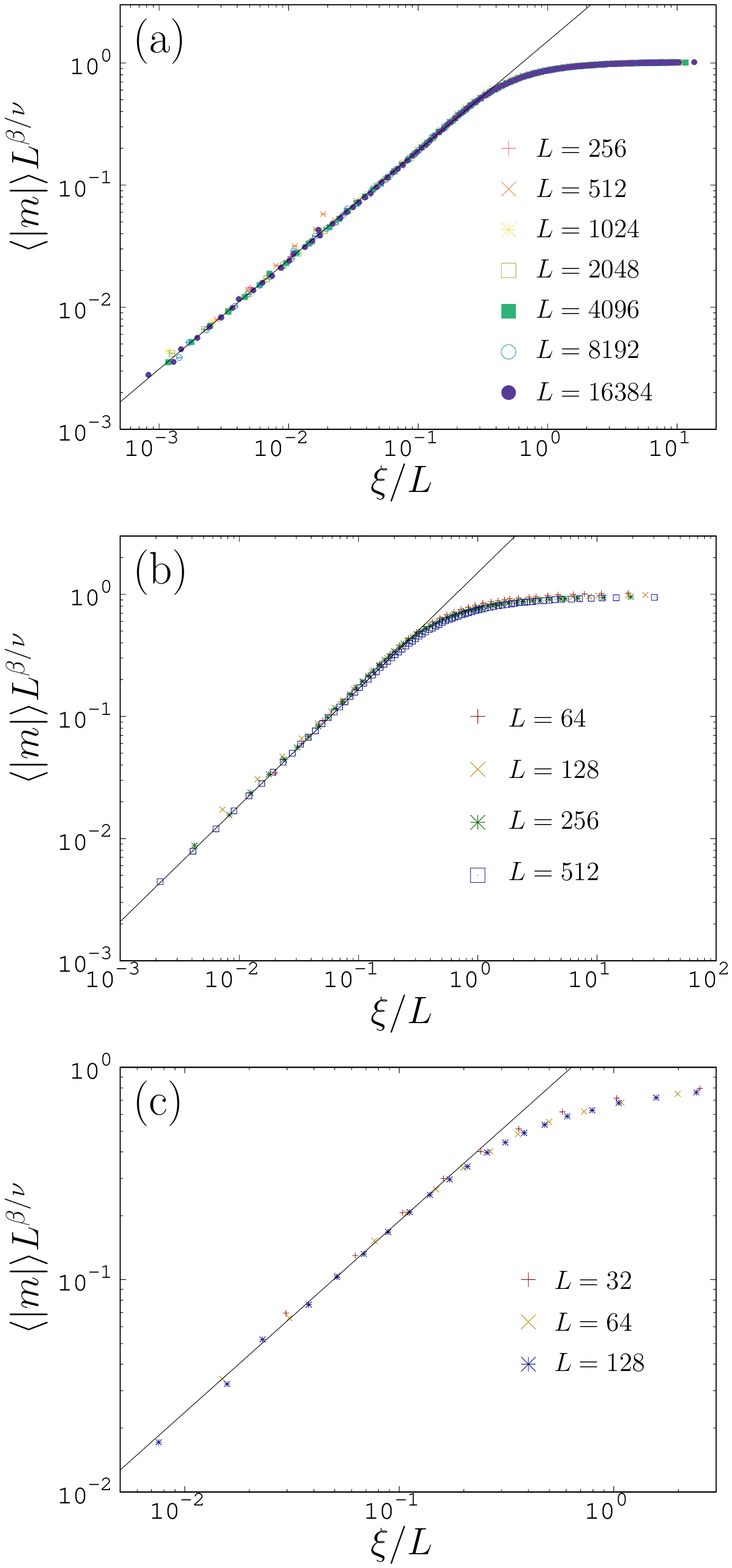}
  \caption{\label{fig:magxi}
    Finite-size scaling plots of the magnetization for (a) the two-, (b) three-, and (c) four-dimensional Ising models.
    Lines show fitting results for the scaled magnetizations
    as functions of the power of $\xi(t)/L$.
  }
\end{figure}

The procedure is as follows:
(i) As in the usual FSS analysis of equilibrium simulations, 
the transition temperature is estimated at first.
The correlation length as a function of Monte Carlo step $t$
converges to certain values at off critical temperatures,
while it constantly develops to the order of the system size
at the critical temperature.
(ii) After the critical temperature is estimated,
critical exponents are estimated by adjusting the vertical
scaling factor, $L^{\omega/\nu}$,
for a quantity in the nonequilibrium relaxation
as shown in Fig.~\ref{fig:magxi}.
Replacing the Monte Carlo step $t$ by the normalized correlation length
$\xi(t)/L$ as the FSS variable,
it is no longer need to estimate the dynamical exponent $z$,
which is indispensable for the estimation of critical exponents
in the NER method~\cite{OzekiIto2007}.

The essence of the method is to describe physical quantities
by a dimensionless variable like the normalized correlation length.
The normalized correlation length directly reflects
correlations of the system, and it helps us to gain insights
into relations between physical quantities and the correlation length.
In the paper, the normalized correlation length is chosen as
the dimensionless variable.
However, any dimensionless variables can be used
to suit various purposes.
For example, the Binder ratio~\cite{Binder1981} is
an alternative dimensionless variable for the cluster NER method~\cite{NonomuraTomita2018}.
The Binder ratio can be utilized as an indicator of
the distribution of order parameter,
which takes a trivial value at a trivial fixed point
and a non-trivial value at a non-trivial fixed point.
Since the Binder ratio is composed of moments of
macroscopic order parameters,
the analysis using the Binder ratio could be efficient
for systems which exhibit non-uniformly ordered phase.

\section{%
  \label{sec:discussion}
  Summary and Discussion
}

The nonequilibrium relaxations at the critical points with the cluster-update
in the Ising models are examined in view of the development
of clusters.
The analytic form of the nonequilibrium relaxation function
of the IR Ising model clarifies that
the relaxation is described by
the product of the power and simple exponential function.
The NER is described by the power-law relaxation
at the very beginning,
and it shows the simple exponential relaxation
just before reaching to the thermally equilibrium state.
Furthermore, using the peculiarity of the complete graph,
it is shown that the system size dependence of
the cluster-flip dynamics is ballistic.
These features in the IR Ising model are different from
the results observed in the two- and three-dimensional Ising models,
in which the relaxation is described by the stretched-exponential form
and the system size dependence of the dynamics is not the usual power-law.
At the four-dimension, which is the upper critical dimension
of the Ising model,
the system exhibits
the power-law relaxation at the very beginning,
and the finite-size time scale is proportional to $L^z$.
These features are equivalent to those of the IR Ising model,
while the dynamical exponent $z$ is smaller than that of the IR Ising model.
This difference in $z$ would come from the finite-size correction.
We expect that $z$ is unity in the four-dimension
since the fluctuation is suppressed enough in the upper critical dimension
and the merging rate described in Eq.~(\ref{eq:mr_irising}),
which deduces $z=1$,
would also be realized in the four-dimensional system.
In addition, at the upper critical dimension,
the estimation of the critical exponents is difficult
because the logarithmic correction severely affects the estimation.
There is another possibility that
the dynamical exponent depends on the dimensionality
and that it converges to unity in the limit of $d\to\infty$.
The examination of the dimensional dependence of the dynamical exponent
is left for future studies.

As the finite-size dependence of the time scale depends on
the dimensionality,
the relaxational dynamics also depends on the dimensionality.
While the stretched-exponential relaxation is observed at $d=2$ and $3$,
the nonequilibrium relaxation is described by
the product of the power and simple exponential function
at $d=4$ and $d=\infty$ (the IR Ising model).
Therefore, there is a dynamical transition point between $d=3$ and $4$.
Even though they are rough analyses, Eqs.~(\ref{eq:diff_xi})
and (\ref{eq:merging_rate}) grasp the mechanism of
cluster growth.
In the IR Ising model, all the sites reside on the surface of
clusters, and the major cluster is able to merge minor clusters
at every sites.
The ease of merging brings about the rapid decrease of minor clusters
and of the merging rate.
On the other hand, in the finite-dimensional Ising models,
only the sites on the surface are able to merge minor clusters.
The restriction of the merging process moderates
the decrease of the merging rate, and it causes
the dynamical transition depending on the dimensionality.
The merging rate described in Eq.~(\ref{eq:merging_rate}) could be
too simplified, but it indicates that the dimensionality of the system
and clusters' surface contribute to the essential feature
of the cluster-flip dynamics.
According to the droplet theory~\cite{HuseFisher1987},
it is known that the decay of the temporal autocorrelation function
in the ordered phase is stretched-exponential for $d < 3$,
while it is simple exponential decay for $d > 3$.
Although the cluster-flip dynamics in the nonequilibrium state
is different from those dealt with the droplet theory,
our results imply the droplet theory would also be effective
for analyzing the cluster-flip dynamics.
Further investigation is required to understand
the universality class of the cluster-flip dynamics.

A new method for the investigation of the critical phenomena
with the nonequilibrium cluster-flip update is described in Sec.~\ref{sec:cner}.
The applicability of the method is restricted to
systems in which the cluster-flip update is effective.
However, the method does not require to equilibrate systems,
and an acceleration of relaxation by the cluster-flip update
reduces computational time significantly.
Combination use of Kibble-Zurek scaling ansatz
will deepen our understanding of the cluster-flip dynamics.

While relaxational dynamics with the single-spin-flip update are
intensively studied, studies of those with the cluster-flip update are
not necessarily enough.
One of the reasons for being overlooked is that
the cluster-flip update is merely seemed to be useful
but its dynamics are thought to be non-physical.
However, our results show that its dynamics reflect states of systems,
and extracting features of systems is possible
by analyzing nonequilibrium cluster-flip relaxations.
There still remains much to be clarified in the cluster-flip dynamics,
and further investigations will develop computational methods
and deepen an understanding of dynamics in spin systems.

\begin{acknowledgments}
  This work was supported by JSPS KAKENHI Grant Number 16K05493.
\end{acknowledgments}

\onecolumngrid

\appendix
\section{%
  \label{appendix:a}
  Magnetization of the Infinite-Range Ising Model near the Critical Point
}

Using the Hubbard-Stratonovich transformation,
the partition function of the IR Ising model [Eq.~(\ref{eq:irmodel})]
is given by~\cite{NishimoriOrtiz2011}
\begin{equation}
  Z = e^{K/2}\sqrt{\frac{NK}{2\pi}}\int_{-\infty}^{\infty}\exp\{-Ng(K; m)\}dm,
\end{equation}
where $g(K; m)$ is
\begin{equation}
  g(K; m) = \frac{K}{2}m^2 - \ln(2\cosh Km).
\end{equation}
Here $N$ is the number of sites, and $K(=\beta J)$ is the exchange interaction
multiplied by the inverse temperature.
Assuming $m \ll 1$, we expand $g(K; m)$ up to the fourth-order in $m$
and obtain
\begin{equation}
  g(K; m) \simeq -\ln 2 + \frac{K(1-K)}{2}m^2 + \frac{K^4}{12}m^4.
\end{equation}
By replacing $NK^4m^4$ by $12\mu^2$, the partition function is written by
\begin{equation}
  Z = \frac{2^{N-1}(12N)^{1/4}}{\sqrt{2\pi K}}
  \exp\left(\frac{K}{2} + \frac{3(1-K)^2N}{4K^2}\right)I,
\end{equation}
where $I$ is
\begin{equation}
  I = \int_0^{\infty}\mu^{-1/2}\exp\left\{
  -\left(\mu + \frac{(1-K)\sqrt{3N}}{2K}\right)^2
  \right\}d\mu.
\end{equation}
This integral $I$ is represented by the summation of gamma functions
by replacing $(\mu + \sqrt{\epsilon})^2$ by $t$,
\begin{align}
  I &= \frac{1}{2}\int_{\epsilon}^{\infty}t^{-1/2}(t^{1/2} - \epsilon^{1/2})^{-1/2}
  e^{-t}dt\nonumber\\
  &= \frac{1}{2}\int_{\epsilon}^{\infty}\sum_{n=0}^{\infty}\frac{(2n-1)!!}{2^nn!}
  t^{-(2n+3)/4}\epsilon^{n/2}dt\nonumber\\
  &= \frac{1}{2}\sum_{n=0}^{\infty}\frac{(2n-1)!!}{2^nn!}
  \left\{\Gamma\left(-\frac{2n-1}{4}\right)
  -\gamma\left(-\frac{2n-1}{4}, \epsilon\right)\right\}\epsilon^{n/2},
\end{align}
where, $\epsilon = 3(1-K)^2N/(4K^2)$,
$\Gamma(a)$ is the gamma function,
and $\gamma(a, x)$ is the lower incomplete gamma function.
The lower incomplete gamma function has the following asymptotic equation~\cite{AbramowitzStegun1964},
\begin{equation}
  \gamma(a, x) = \Gamma(a)x^{a}e^{-x}\sum_{n=0}^{\infty}\frac{x^n}{\Gamma(a+n+1)}.
\end{equation}
Using the asymptotic equation, the integral $I$ is rewritten by
\begin{equation}
  I = \frac{1}{2}\sum_{n=0}^{\infty}\frac{(2n-1)!!}{2^nn!}
  \Gamma\left(-\frac{2n-1}{4}\right)
  \left(\epsilon^{n/2} - \epsilon^{1/4}e^{-\epsilon}
  \sum_{k=0}^{\infty}\frac{\epsilon^k}{\Gamma((5-2n)/4+k)}
  \right).
\end{equation}
Therefore, the partition function up to the fourth-order in $m$ is given by
\begin{equation}
  \label{eq:Z_IR}
  Z = \frac{2^{N-2}(12N)^{1/4}e^{K/2}}{\sqrt{2\pi K}}
  \sum_{n=0}^{\infty}\frac{(2n-1)!!}{2^nn!}\Gamma\left(-\frac{2n-1}{4}\right)
  \left(
  \epsilon^{n/2}e^{\epsilon} - \epsilon^{1/4}\sum_{k=0}^{\infty}\frac{\epsilon^k}{\Gamma((5-2n)/4+k)}
  \right).
\end{equation}
In a similar manner, the thermal average of the magnetization
times the partition function, $\langle m\rangle Z$, can be obtained as
\begin{align}
  \label{eq:mZ_IR}
  \langle m\rangle Z &= e^{K/2}\sqrt{\frac{NK}{2\pi}}\int_{-\infty}^{\infty}
  m\,\exp\{-Ng(K; m)\}dm\nonumber\\
  &= \frac{2^{N-2}\sqrt{12}e^{K/2}}{\sqrt{2\pi K^3}}
  \Gamma\left(\frac{1}{2}\right)
  \left(
  e^{\epsilon} - \epsilon^{1/2}\sum_{k=0}^{\infty}\frac{\epsilon^k}{\Gamma(3/2 + k)}
  \right).
\end{align}
Combining Eq.~(\ref{eq:Z_IR}) and Eq.~(\ref{eq:mZ_IR}),
we obtain the magnetization near the critical point as
\begin{align}
  \label{eq:irmodel_mag}
  \langle m \rangle &=
  \frac{12^{1/4}\Gamma(1/2)}{N^{1/4}K}
  \left(
  e^{\epsilon} - \epsilon^{1/2}\sum_{k=0}^{\infty}\frac{\epsilon^k}{\Gamma(3/2 + k)}
  \right)
  \left\{
  \sum_{n=0}^{\infty}\frac{(2n-1)!!}{2^nn!}\Gamma\left(-\frac{2n-1}{4}\right)
  \right.\nonumber\\
  &\quad
  \left.
  \left(
  \epsilon^{n/2}e^{\epsilon} - \epsilon^{1/4}\sum_{k=0}^{\infty}\frac{\epsilon^k}{\Gamma((5-2n)/4+k)}
  \right)
  \right\}^{-1}.
\end{align}
The Taylor series expansion of $\langle m\rangle$ with respect to
$\epsilon^{1/4}$ is given by
\begin{equation}
  \label{eq:irmodel_mag2}
  \langle m \rangle =
  \frac{12^{1/4}\Gamma(1/2)}{N^{1/4}\Gamma(1/4)K}
  \left(1 + \sum_{n=1}^{\infty}c_n\epsilon^{n/4}\right),
\end{equation}
where $c_n$ is the $n$th Taylor coefficient.
From Eq.~(\ref{eq:irmodel_mag2}),
by setting the variables to the critical values, $K = 1$ and $\epsilon = 0$,
we obtain the value of the magnetization at the critical point,
\begin{equation}
  \label{eq:irmodel_mag_tc}
  \langle m_{\mathrm{C}} \rangle =
  \frac{12^{1/4}\Gamma(1/2)}{N^{1/4}\Gamma(1/4)}.
\end{equation}

\twocolumngrid

\section{%
  \label{appendix:b}
  Nonequilibrium relaxation of the magnetization in the IR Ising model
}

Using the peculiarity of the complete graph,
we derive the development of magnetization of the infinite-range Ising model
at the critical point as a function of Monte Carlo step.
The magnetization depends on the density of bonds,
which are fundamental elements in the graph representation,
and we consider the time evolution of the density of bonds
rather than the magnetization directly.
In the Swendsen-Wang cluster algorithm,
we put bonds between parallel spins with the probability $p$.
Denoting the magnetization at Monte Carlo step $n-1$ by $M_{n-1}$,
the number of spins parallel to the magnetization $N_{+}$
and the number of spins antiparallel to the magnetization $N_{-}$,
respectively, is given by
\begin{align}
  N_{+} &= \frac{N + M_{n-1}}{2},\\
  N_{-} &= \frac{N - M_{n-1}}{2}.
\end{align}
The number of bonds between parallel spins $B_{\parallel}$ is
\begin{equation}
  B_{\parallel} = \frac{N_{+}(N_{+}-1)}{2} + \frac{N_{-}(N_{-}-1)}{2}.
\end{equation}
The probability of a putting bond is
\begin{equation}
  p = 1 - e^{-2K/N}.
\end{equation}
For large enough $N$,
the probability $p$ near the critical point ($K\sim 1$)
is approximately represented as
\begin{equation}
  p \simeq \frac{2}{N}.
\end{equation}
The density of bonds at Monte Carlo step $n$, $b_n$, is
given by
\begin{align}
  b_n &= \frac{B_{\parallel}}{B}p,\nonumber\\
  \label{eq:recurr0}
  &= \frac{1}{N} + \frac{1}{N}m^2_{n-1},
\end{align}
where $B$ is the number of bonds and $m_n$ is the magnetization
per site at Monte Carlo step $n$.
Near the critical point, the magnetization per site $m$
(see Eq.~(\ref{eq:irmodel_mag2}))
is given by 
\begin{equation}
  \label{eq:mag}
  m(x) = AN^{-1/4}(1 - \sum_{i=1}^{\infty}C_ie^{-3x^4/4}x^i),
\end{equation}
where $A$ is a constant, $C_i$ is a expansion coefficient,
and $x = [(1-K)/K]^{1/2}N^{1/4}$.
In the quench process, we assume $x \ll 1$ and $(1-K)/K \propto b_c - b$,
where $b_c$ is the density of bonds at the critical point,
and $b$ is the density of bonds.
Considering the slowest term to converge,
Eq.~(\ref{eq:recurr0}) is rewritten by
\begin{align}
  b_n &= \frac{1}{N} + \frac{A_1}{N^{3/2}}\{1 - A_2N^{3/2}(b_c - b_{n-1})\},\\
  \label{eq:recurr}
  &\simeq \frac{1}{N} - A_1A_2(b_c - b_{n-1}).
\end{align}
Here, $A_1$ and $A_2$ are constants.
Solution of this recurrence relation (Eq.~(\ref{eq:recurr})) is given by
\begin{equation}
  \label{eq:bn}
  b_n = b_{\infty} - (b_{\infty} - b_0)r^n,
\end{equation}
where $r = A_1A_2$ and $b_{\infty} = (1/N-rb_c)/(1-r)$.
Combining Eq.~(\ref{eq:bn}) and Eq.~(\ref{eq:mag}),
the magnetization at Monte Carlo step $n$, $m_n$, is given by
\begin{equation}
  \label{eq:mn0}
  m_n \simeq AN^{-1/4}[1 - C\{b_c - b_{\infty} + (b_{\infty} - b_0)r^n\}^{1/2}N^{3/4}].
\end{equation}
In the critical region, inside of the curly bracket
is proportional to $N^{-3/2}$, and we obtain
\begin{equation}
  \label{eq:mn}
  m_n \simeq AN^{-1/4}[1 - c_1(1 + c_2e^{-n/\tau})^{1/2}],
\end{equation}
where $c_1$ and $c_2$ are constants.

\bibliography{tomita}

\end{document}